# Enabling Down Syndrome Research through a Knowledge Graph–Driven Analytical Framework

Madan Krishnamurthy[1], Surya Saha[2], Pierrette Lo[3], Patricia L. Whetzel[1], Tursynay Issabekova[1], Jamed Ferreris Vargas[4], Jack DiGiovanna[2], Melissa A Haendel[1]

[1]University of North Carolina at Chapel Hill, Chapel Hill, NC, USA

[2]Velsera, Charlestown, MA, USA

[3]Linda Crnic Institute for Down Syndrome, University of Colorado Anschutz Medical Campus, Aurora, CO, USA

[4]Vanderbilt University Medical Center, Nashville, TN, USA

**ABSTRACT**

**Background:** Trisomy 21 results in Down syndrome (DS), a multifaceted genetic disorder manifesting through varied clinical phenotypes such as structural heart anomalies, immune system irregularities, intellectual disabilities, and early-onset dementia risk. The clinical diversity, combined with fragmented data across multiple studies, creates significant challenges for comprehensive research and translational discovery. The NIH INCLUDE (INvestigation of Co-occurring conditions across the Lifespan to Understand Down syndromE) initiative has assembled a robust repository of harmonized participant-level data, but fully realizing its potential requires advanced analytical frameworks that enable cross-study integration and AI-driven discovery.

**Methods:** We developed a knowledge graph-driven platform that transforms nine individual INCLUDE studies (comprising 7,148 participants, 456 conditions, 501 phenotypes, and over 37,000 biospecimens) into a unified semantic infrastructure. Our approach combines semantic integration using domain-aware RDF schemas with cross-resource enrichment from the Monarch Initiative, expanding entity coverage to include 4,281 genes and 7,077 variants alongside original clinical data.

**Results:** The resulting knowledge graph contains over 1.6 million semantic associations, enabling AI-ready analysis through graph embeddings and path-based reasoning for hypothesis generation. SPARQL querying and natural language interfaces provide intuitive access for researchers, while graph analysis revealed 79 shared phenotypes across JAK-STAT pathway genes.

**Conclusions:** This framework transforms static data repositories into dynamic discovery environments, enabling systematic exploration of genotype-phenotype relationships, cross-study pattern recognition, and predictive modeling to advance understanding and care for individuals with Down syndrome.

**INTRODUCTION**

Down syndrome (DS) is a complex, multisystem neurodevelopmental condition resulting from an extra copy of chromosome 21 (trisomy 21)[1]. Individuals with DS experience a wide range of health challenges, including congenital heart defects, immune and endocrine dysfunction, neurodevelopmental differences, and an elevated risk of early-onset Alzheimer's disease[2]. This clinical heterogeneity presents challenges for diagnosis, treatment, and longitudinal research, highlighting the importance of integrated, cross-domain approaches that support insights across the lifespan.

The NIH INCLUDE (INvestigation of Co-occurring conditions across the Lifespan to Understand Down syndromE)[3] initiative has made significant strides in this direction by assembling a robust, multi-study repository of harmonized participant-level data. This includes demographics, biospecimens, clinical conditions, phenotypic traits, and genomic profiles forming a vital foundation for discovery and translational impact. These diverse datasets form the foundational **data** layer. Through harmonization, semantic annotation, and integration across multiple studies, this data is transformed into **information** that is interpretable and interoperable. Harmonization specifically refers to transforming heterogeneous study data into the INCLUDE Data Coordinating Center's[4] LinkML[5]-based common data model[6], ensuring consistency in variables, terminologies, and formats across studies. By constructing a Knowledge Graph[7] (KG) enriched with curated gene–disease–phenotype relationships, this information is further elevated into **knowledge** capable of driving hypothesis generation, predictive modeling, and translational insights, reflecting the DIKW (Data–Information–Knowledge–Wisdom)[8] paradigm in biomedical research.

To fully harness the potential of this rich resource, we introduce a KG–driven analytical platform that enables deeper integration, semantic enrichment, and AI-ready infrastructure. Our approach builds upon the INCLUDE data assets to support scalable, cross-study exploration, knowledge-driven hypothesis generation, and predictive modeling.

Key capabilities of this platform include:

- **Semantic integration** of structured datasets into a unified KG[7] using domain-aware schema models;
- **Cross-resource enrichment** with curated gene–disease–phenotype knowledge from sources such as the Monarch Initiative[9];
- **AI-readiness** through the use of graph embeddings[10] that enable predictive modeling, link prediction, and similarity-based analysis;
- **Path-based querying and reasoning**, allowing users to explore multi-hop biological and clinical relationships (e.g., Participant → Condition → Gene → Drug);
- **Natural language interfaces**, enabling clinicians and researchers to pose complex questions through LLM[11]-to-SPARQL[12] translation.

By transforming the INCLUDE resource into a dynamic knowledge environment, this framework opens new avenues for discovery, allowing researchers to uncover latent patterns, explore connections across domains, and build tailored predictive models to advance understanding and care for individuals with DS. Specifically, the platform makes available both machine-readable RDF KGs and reproducible analysis workflows (e.g., scripts, notebooks).

**METHODS**

As illustrated in **Figure 1**, our framework integrates harmonized participant-level datasets from multiple secure repositories, including Synapse[13], AWS S3[14] buckets, and the project portal[15], into a semantically rich, queryable KG. Our KG transforms nine individual NIH INCLUDE studies[15] **HTP**, **X01-Hakonarson**, **X01-deSmith**, **BRI-DSR**, **DSC**, **DS-Sleep**, **ABC-DS**, **TEAM-DS**, and the **ALL (merged)** dataset into an integrated semantic infrastructure. Each study contributes distinct data modalities, ranging from phenotypic and biospecimen records to neuroimaging and genetic assays. The merged ALL KG captures the integrated scope, enabling cross-study exploration and inference. In the **knowledge representation**[16,17] phase, harmonized data are transformed into structured graph entities using well-defined ontologies and controlled vocabularies. This process includes both *knowledge generation*, which encodes the data into the KG structure, and *knowledge enrichment*, which links entities to external databases and semantic resources to enhance coverage and interoperability. The resulting KG serves as the foundation for **knowledge discovery**, where analytical methods such as *graph analysis* reveal patterns and relationships, while *graph embeddings* enable machine learning applications including clustering, classification, and link prediction. Finally, the framework supports **knowledge exploration** through *SPARQL* queries, interactive *visualizations*, and a natural language *chatbot* interface, providing researchers with intuitive, multi-modal access for data interrogation, hypothesis generation, and insight derivation.

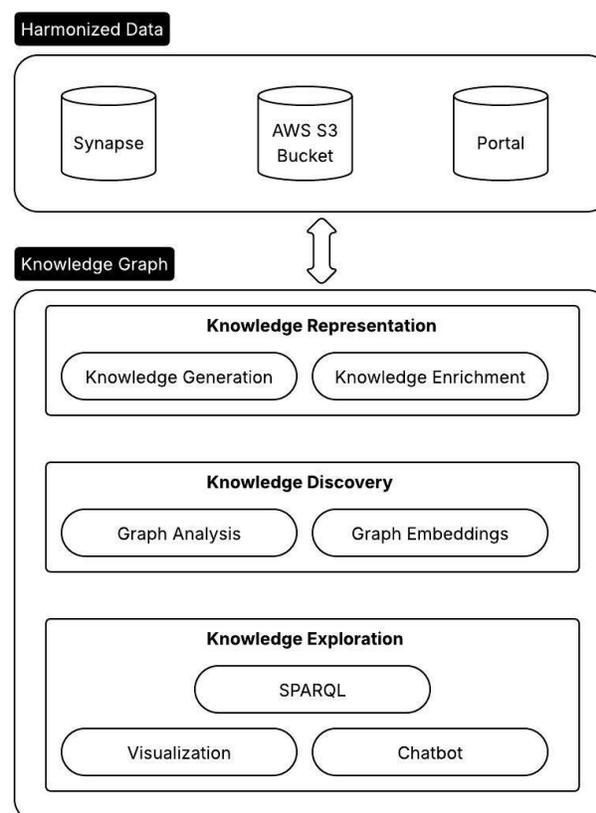

FIGURE 1. Framework for Transforming and Exploring NIH INCLUDE Study Data via a Knowledge Graph

*KNOWLEDGE GENERATION*

The **knowledge generation** phase involved transforming harmonized participant-level data from the NIH INCLUDE Data Coordinating Center (DCC)[18] into a semantically rich, queryable KG. This process is built directly upon the **INCLUDE LinkML data model**[6], which defines the classes, attributes, and controlled vocabularies used to describe studies, participants, conditions, phenotypes, biospecimens, and related entities.

Using **rdflib**[19], **domain-aware RDF schema**[20] was derived from this model, preserving the semantics and structure of the original specification while explicitly representing relationships in a graphical form. Relationships such as `hasParticipant`, `hasCondition`, `hasPhenotype`, `hasBiospecimen`, `hasDataFile`, and `hasMedicalAction` were defined to capture connections between entities and enable both biological and clinical queries.

All original model features were retained as **annotation properties** on the corresponding RDF classes and instances. This ensured that descriptive and provenance metadata including study design, participant demographics, biospecimen collection parameters, phenotypic descriptors, and assay metadata remained accessible within the KG without loss of fidelity to the original schema.

**Schema**

The schema (**Figure 2**) captures the primary entity classes and their interrelationships, representing the logical organization of INCLUDE's multi-modal data. By explicitly defining these links, the schema supports traversal across domains (e.g., `Study → Participant → Condition → Phenotype → Gene → Drug`) and enables integration with external biomedical ontologies[21] such as MONDO[22], HPO[23].

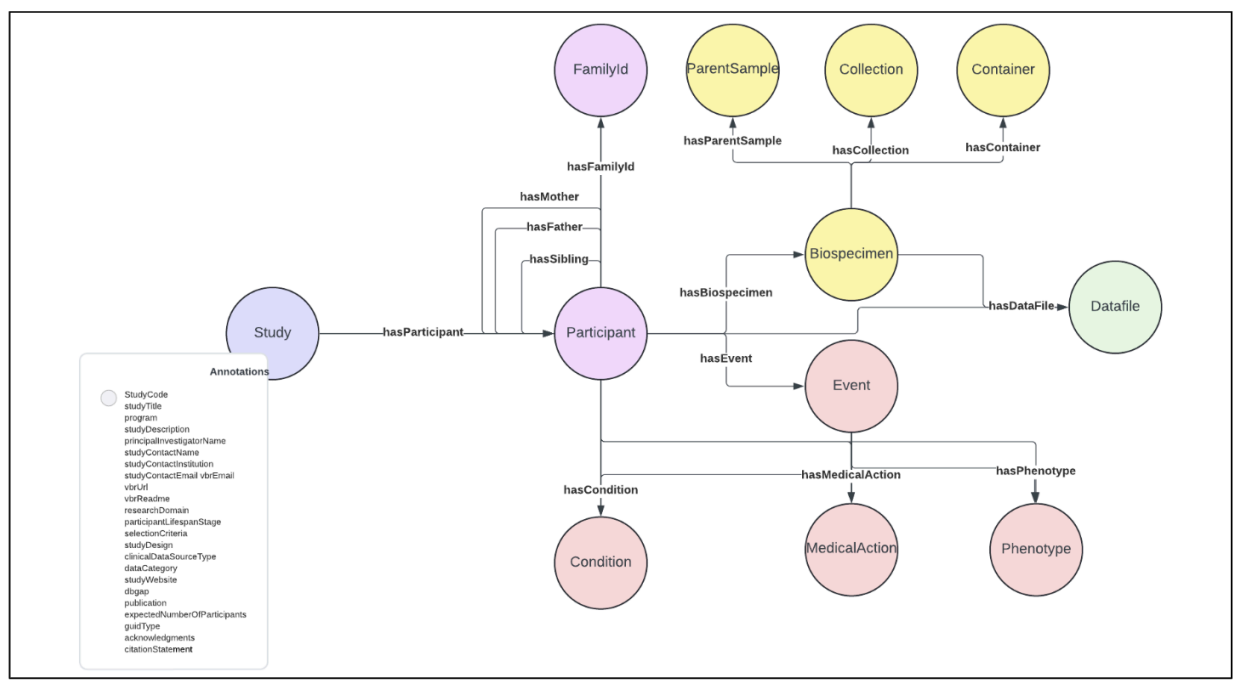

**FIGURE 2. INCLUDE Knowledge Graph Schema**

**Instance**

Data ingestion scripts were implemented to map CSV-based harmonized datasets into RDF triples according to the schema. Separate loaders were developed for each entity type:

- **Study loader** – Parses and instantiates Study nodes with design, funding, and thematic metadata.
- **Participant loader** – Creates Participant nodes linked to their parent Study and annotated with demographic, familial, and clinical attributes.
- **Event loader** – Generates Event instances connected to participants through **hasEvent**, encapsulating conditions, phenotypes, and medical actions, along with age-at-observation and event identifiers.
- **Biospecimen loader** – Represents sample metadata, collection events, containers, and parent–child sample relationships.
- **DataFile loader** – Connects biospecimens and participants to associated data files, including access URIs, formats, and experimental strategy descriptors.

All instance URIs were generated using a **normalized URI creation function** to ensure identifier stability and interoperability. RDF serialization in **Turtle format**[24] was used for persistence and downstream querying.

By unifying the INCLUDE data under a common schema and preserving rich annotations, the resulting FAIR-compliant (Findable, Accessible, Interoperable, Reusable)[25] KG forms a robust foundation for scalable, semantically informed analyses across studies and modalities. This unified framework supports cross-cutting queries, enables discovery of phenotypic patterns across cohorts, facilitates tracing of biospecimen-linked genomic datasets for specific disease subtypes, and empowers knowledge inference, cohort discovery, and seamless integration with federated biomedical data networks.

*KNOWLEDGE ENRICHMENT*

The initial INCLUDE KG, derived from harmonized participant-level data in the NIH INCLUDE Data Coordinating Center (DCC), captured a defined set of biomedical entities such as diseases (MONDO[22]) and phenotypes (HPO[23]) and their relationships as represented in the source datasets. Since the harmonized data model already incorporated ontology-based annotations (e.g., MONDO for diseases and HPO for phenotypes), these terms were directly leveraged in the KG to enable consistent integration with external biomedical ontologies. While valuable, this primary KG was inherently incomplete, constrained by the scope and coverage of the originating data. Many clinically relevant associations, mechanistic links, and intermediate concepts were absent, limiting both translational insight and computational reasoning potential.

**Rationale for Enrichment**

Biomedical knowledge is inherently distributed across numerous heterogeneous resources. No single dataset offers complete coverage of disease mechanisms, genotype–phenotype correlations, or variant impacts. Without augmentation, KGs risk under-representing biologically important relationships, thereby restricting cross-domain connectivity and multi-hop inference.

Knowledge enrichment addresses this gap by systematically integrating curated associations from external, authoritative resources to increase connectivity, improve concept coverage, and enhance the diversity of semantic relationships.

**Targeted Growth Strategy**
 We implemented a directed enrichment workflow that started with four core entity classes from the harmonized INCLUDE dataset—Conditions (*MONDO*[22]), Phenotypes (*HPO*[23]), Genes (*HGNC*[26]), and Variants (*ClinVar*[27]). To prevent combinatorial explosion while maximizing biological relevance, we applied class-specific growth rules (**Figure 3**):

- **Condition enrichment** retrieved only phenotype, gene, and variant associations.
- **Phenotype enrichment** retrieved only condition, gene, and variant associations.
- **Gene enrichment** retrieved only condition, phenotype, and variant associations.
- **Variant enrichment** retrieved only condition, phenotype, and gene associations.

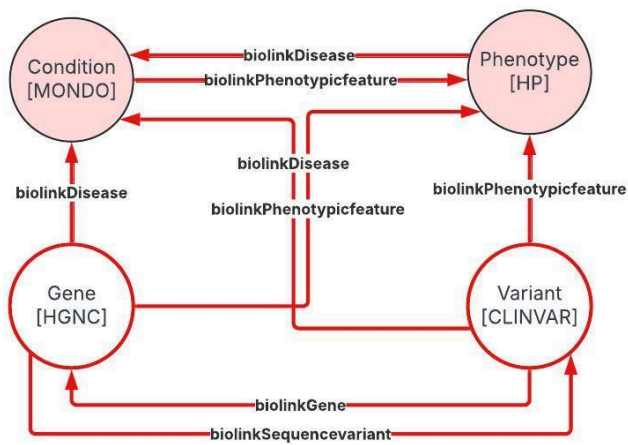

FIGURE 3. Enrichment of the INCLUDE Knowledge Graph with External Biomedical Resources

This ensured that enrichment expanded cross-domain links without generating redundant same-class associations (e.g., disease–disease).

**Monarch Integration as a Use Case**
 We selected the Monarch Initiative KG as the primary enrichment source due to its aggregation of biomedical associations from multiple curated databases (e.g., ClinVar, OMIM[28], Orphanet[29], HPO annotations) and its alignment to the Biolink Model[30] semantic standard. For each seed entity, the Monarch API[31] was queried in both subject and object roles to capture bidirectional associations. Associations were filtered to remove:

1. Hierarchical relations (`biolink:subclass_of`),
2. Same-class entity links, and
3. Non-standard identifiers incompatible with INCLUDE ontologies.

Validated associations were integrated into the KG with CURIE resolution, ontology-consistent typing, human-readable labels, Biolink-compliant predicates, and provenance annotations (`sourceAnnotation="monarch"`).

**Outcome and Extensibility**
 This iterative enrichment continued until all eligible seeds were processed, yielding substantial expansion in entity coverage and cross-domain connectivity. The approach increased semantic density, enabled new multi-hop inference paths (e.g., *Condition → Gene → Variant → Phenotype*), and preserved ontological precision. While Monarch served as the exemplar data source, the methodology is resource-agnostic and can incorporate other structured biomedical knowledge bases (e.g., DisGeNET[32], GWAS Catalog[33], REACTOME[34], STRING[35]) without altering the core framework—further extending the translational and clinical applicability of the INCLUDE KG.

*KNOWLEDGE DISCOVERY*

**Graph Embedding**
After knowledge generation and enrichment, the harmonized KG was transformed into an AI-ready format to enable advanced knowledge discovery. AI-readiness refers to the extent to which data is structured, standardized, and semantically enriched to support machine learning (ML), deep learning (DL), and other automated analytical approaches[36]. To achieve this, we converted the RDF triples into compact, information-rich numerical representations using a KG embedding workflow implemented in PyKEEN[37].

The process began by exporting subject–predicate–object triples from the RDF graph into a simplified, tabular format with normalized entity identifiers and preserved literal values. These triples were loaded into PyKEEN's `TriplesFactory`, enabling the use of state-of-the-art embedding models. We selected the TransE[38] model with a 250-dimensional embedding space, trained using a margin-based ranking loss under the stochastic local closed-world assumption (sLCWA) sampling strategy. Training was performed for 10 epochs on GPU where available, and model performance was evaluated using rank-based metrics. The resulting embeddings, along with entity-to-ID mappings, were stored for reuse in downstream analyses.

KG embeddings encode the KG's structural and semantic patterns into continuous vector spaces, enabling a wide range of AI-driven discovery tasks. They can support **link prediction**, where novel gene–disease associations are inferred by identifying likely missing edges; **semantic/similarity search**, which finds participants, biospecimens, diseases, or genes with comparable characteristics based on vector proximity; and **clustering**, which groups entities by latent features to reveal hidden subpopulations[39]. Embeddings also facilitate **outlier detection**, allowing researchers to spot disconnected or unusual nodes that may represent data errors or novel biological phenomena. Finally, they can be used for **downstream modeling**, serving as powerful input features for predictive tasks such as classification, regression[40].

In this study, we illustrate the downstream modeling capability by training a classifier to predict DS status directly from participant embeddings. Additionally, to enhance interpretability, we project the high-dimensional embeddings into two dimensions for visualization in the results section, enabling an intuitive understanding of the data's latent structure.

**Graph Analysis**
Following the generation of AI-ready KG embeddings, we performed complementary graph analysis to directly interrogate the semantic structure of the NIH INCLUDE integrated KG. Graph analysis enables us to explore explicit, ontology-driven connections between entities without relying solely on latent representation learning. While the framework supports a wide range of network analysis tasks using libraries such as NetworkX[41]—e.g., degree distribution profiling, centrality analysis, and community detection—in this study we focus on a path-based use case to illustrate targeted exploration of biological relationships.

In the path-based analysis[42], we implement a breadth-first search[43] (BFS) traversal over the RDF graph to identify and summarize semantically valid paths connecting a given start entity to participants. Path traversal is constrained by an ontology-informed whitelist of predicates (e.g., *hasPhenotype*, *biolink:Gene*, *biolink:Disease*) and by excluding certain entity types (e.g., *Event*) to ensure biologically meaningful connectivity. This approach enables the discovery of intermediate entities—such as phenotypic features, conditions, or variants—that link a specific gene or disease to subsets of participants across studies.

As a case study, we define cohorts based on genetic features by initiating BFS from curated gene identifiers (e.g., [HGNC:6190](), [HGNC:3091]()) and collecting all reachable participants along allowed semantic paths. The traversal results are aggregated into path summaries, which are then visualized as Sankey diagrams[44] to convey multi-hop relationships between genes, intermediate biomedical entities, and participant cohorts. We further compute an entity presence matrix across multiple start nodes to identify common phenotypes or conditions shared across gene-defined subpopulations.

This path-based analysis demonstrates how the integrated KG can support fine-grained, ontology-driven cohort definition and hypkaothesis generation. It complements the embedding-based approaches by providing interpretable, explicit relationship chains that can be traced back to their data sources—facilitating both exploratory analysis and the validation of machine learning–derived patterns.

*KNOWLEDGE EXPLORATION*

The final phase of the framework focuses on **interactive, multi-modal access** to the NIH INCLUDE KG for hypothesis generation, cohort definition, and exploratory analysis. This phase integrates **ontology-driven SPARQL querying** with a **natural language chatbot interface**, allowing both programmatic and conversational interrogation of the KG.

**SPARQL Querying**
Using the **rdflib** library in Python, we implemented a suite of SPARQL queries for targeted retrieval of biomedical entities and their relationships across studies. Example queries include:

- Aggregating **conditions** (MONDO terms) by participant counts and identifying those present across multiple studies.
- Retrieving **phenotypes** (HPO terms) observed in ≥5 studies, enabling cross-study phenotype harmonization.
- Linking **participants** to conditions, phenotypes, biospecimens, and data files to support downstream cohort assembly.

Results are returned in tabular form and can be further processed into domain-specific summaries, entity co-occurrence matrices, and visualizations such as UpSet[45] plots for intersection analysis. These queries leverage the explicit ontology-aligned predicates in the KG (e.g., `hasCondition`, `hasPhenotype`, `hasBiospecimen`) to ensure semantic consistency and reproducibility.

**Conversational Exploration via Chatbot.**
To make KG exploration accessible to non-technical users, we developed a **Streamlit-based chatbot**[46] **interface** powered by the OpenAI API. The chatbot accepts natural language questions (e.g., *"Which phenotypes are common across more than five studies?"*) and internally maps them to pre-defined or dynamically generated SPARQL queries. Query execution results are formatted into concise, human-readable outputs that can be iteratively refined through follow-up prompts.

The chatbot supports:

- On-demand retrieval of study- and participant-level metadata.
- Entity-specific lookups (e.g., all genes associated with a given phenotype).
- Multi-hop traversals to uncover indirect relationships between biomedical concepts.

While the current chatbot serves as a **proof-of-concept**, it will be enhanced in future versions to improve accuracy, reduce latency, broaden supported query types, and incorporate context-aware reasoning for more sophisticated dialogue-driven exploration.

**Integration and Usability**
This dual approach—precise SPARQL querying for structured analysis and a natural language interface for exploratory dialogue—provides complementary modes of KG access. Researchers can rapidly prototype formal queries, validate findings through explicit graph traversal, and perform open-ended exploration without requiring SPARQL expertise. Together, these capabilities transform the INCLUDE KG from a static data integration product into a dynamic, researcher-friendly environment for **data interrogation, hypothesis generation, and discovery**.

**RESULTS**

*KNOWLEDGE GENERATION*

The knowledge generation pipeline produced fully instantiated INCLUDE KGs for nine individual NIH INCLUDE studies—**HTP, X01-Hakonarson, X01-deSmith, DS-Sleep, ABC-DS, TEAM-DS, BrainPower, DSC, and BRI-DSR**—as well as a merged **ALL** KG that integrates all cohorts into a unified semantic network.

**Entity Instances**
Each KG contains instances of the core schema classes defined in the INCLUDE model—*Study, Participant, FamilyId, Event, Condition, Phenotype, MedicalAction, Biospecimen, Container, ParentSample,* and *Collection*—as summarized in **Table 1**. The distribution of instances reflects study-specific differences in cohort size, clinical scope, and data availability. For example, **HTP** includes over 1,000 participants and more than 14,000 biospecimens, whereas **DS-Sleep** contains participant and event data but no biospecimen records. The merged **ALL** KG aggregates 7,148 participants, 6,962 events, 456 conditions, 501 phenotypes, and over 37,000 biospecimen container entries into a single interoperable graph.

| *Schema/Class* | *Instances* | | | | | | | | | |
|---|---|---|---|---|---|---|---|---|---|---|
| | HTP | X01-Hakonarson | X01-deSmith | DS-Sleep | ABC-DS | TEAM-DS | BrainPower | DSC | BRI-DSR | ALL |
| *Study* | 1 | 1 | 1 | 1 | 1 | 1 | 1 | 1 | 1 | 9 |
| *Participant* | 1055 | 1152 | 436 | 76 | 419 | 126 | 82 | 3634 | 168 | 7148 |
| *FamilyId* | 812 | 691 | | 35 | 41 | 2 | | 27 | 1 | 1609 |
| *Event* | 917 | 1152 | 436 | 36 | 416 | 122 | 82 | 3634 | 167 | 6962 |
| *Condition* | 147 | 269 | 15 | 2 | 55 | 108 | 13 | 111 | 11 | 456 |
| *Phenotype* | 152 | 301 | 16 | | 50 | 129 | 13 | 113 | 7 | 501 |
| *MedicalAction* | 8 | 3 | | | 18 | 27 | | 7 | | 47 |
| *Biospecimen* | 14515 | 1513 | 436 | | | | | | 46 | 16510 |
| *Container* | 37169 | | | | | | | | 23 | 37192 |
| *ParentSample* | 4399 | | | | | | | | 23 | 4422 |
| *Collection* | 2282 | | | | | | | | 23 | 2305 |
| *DataFile* | 11302 | 10224 | 2616 | | | | | | 254 | 24396 |

**TABLE 1. Entity Instance Counts per Study**

**RDF File Generation**

Each study-specific KG was serialized in **Turtle** format and is stored in Synapse with complete provenance metadata linking it to the harmonized source datasets, loader scripts, and transformation parameters (**Table 2**). File sizes range from 69 KB (*DS-Sleep*, 1,626 associations) to 22.9 MB (*HTP*, 486,776 associations). The merged **ALL** KG comprises 54.4 MB and contains more than 1.27 million associations, representing the integrated semantic structure across all studies.

| *Study* | *KG name* | *KG Size* | *# associations* |
|---|---|---|---|
| HTP | INCLUDE_KG_HTP.rdf | 22.9 MB | 486776 |
| X01-Hakonarson | INCLUDE_KG_X01-Hakonarson.rdf | 12.5 MB | 235369 |
| X01-deSmith | INCLUDE_KG_X01-deSmith.rdf | 3.4 MB | 69352 |
| BRI-DSR | INCLUDE_KG_BRI-DSR.rdf | 451 KB | 7710 |
| DSC | INCLUDE_KG_DSC.rdf | 13.2 MB | 403953 |
| DS-Sleep | INCLUDE_KG_DS-Sleep.rdf | 69 KB | 1626 |
| ABC-DS | INCLUDE_KG_ABC-DS.rdf | 1.2 MB | 39418 |
| TEAM-DS | INCLUDE_KG_TEAM-DS.rdf | 600 KB | 20405 |
| BrainPower | INCLUDE_KG_BrainPower.rdf | 291 KB | 8018 |
| ALL (merged) | INCLUDE_ALL.rdf | 54.4 MB | 1270295 |

**TABLE 2. RDF Knowledge Graph Files and Sizes**

**Provenance and Reproducibility**

All RDF outputs are accompanied by detailed provenance records, enabling direct traceability from any KG triple back to its source harmonized file. This ensures full reproducibility, supports transparent auditing of the KG construction process, and allows selective regeneration of KGs as source datasets evolve.

Together, the instance counts and RDF file statistics demonstrate the pipeline's ability to transform heterogeneous, multi-modal cohort datasets into ontology-linked, FAIR-compliant KGs at scale while preserving detailed provenance for reuse and integration.

*KNOWLEDGE ENRICHMENT*

The knowledge enrichment process was applied to the initial INCLUDE KGs for nine individual NIH INCLUDE studies—HTP, X01-Hakonarson, X01-deSmith, ABC-DS, TEAM-DS, BrainPower, DSC, BRI-DSR—and the merged ALL KG. This process substantially expanded the coverage of core biomedical entity classes, including Conditions (MONDO), Phenotypes (HPO), Genes (HGNC), and Variants (ClinVar), as well as their cross-domain associations.

Across all studies, enrichment yielded significant increases in the number of represented entities (see **Table 3**). For example, while the initial KGs contained only Conditions and Phenotypes from the harmonized INCLUDE datasets, enrichment introduced thousands of new gene and variant nodes and dramatically expanded disease–phenotype, disease–gene, and phenotype–gene links. The ALL (merged) KG saw the largest growth in both entity counts and cross-domain connectivity.

Notably, the DS-Sleep study—due to its limited starting set of entities and associations—yielded no measurable enrichment, highlighting the dependency of this approach on the initial KG's scope and coverage.

| Schema/Class | Instances | | | | | | | | |
|---|---|---|---|---|---|---|---|---|---|
| | HTP_MI | X01-Hakonarson_MI | X01-deSmith_MI | ABC-DS_MI | TEAM-DS_MI | BrainPower_MI | DSC_MI | BRI-DSR_MI | ALL |
| | Before -> After (Enrichment) | | | | | | | | |
| Condition | 147 -> 7321 | 269 -> 7826 | 15 -> 2092 | 55 -> 4757 | 108 -> 6747 | 13 -> 2507 | 111 -> 6355 | 11 -> 1921 | 456 -> 9238 |
| Phenotype | 152 -> 9448 | 301 -> 9593 | 16 -> 6234 | 50 -> 8572 | 129 -> 9294 | 13 -> 6966 | 113 -> 9205 | 7 -> 4502 | 501 -> 9856 |
| Gene | 0 -> 3551 | 0 -> 3721 | 0 -> 897 | 0 -> 2179 | 0 -> 3327 | 0 -> 1110 | 0 -> 3133 | 0 -> 566 | 0 -> 4281 |
| Variant | 0 -> 6842 | 0 -> 6865 | 0 -> 2931 | 0 -> 5542 | 0 -> 6191 | 0 -> 4006 | 0 -> 5852 | 0 -> 1332 | 0 -> 7077 |

**TABLE 3. Entity Counts Before and After Knowledge Enrichment**

The resulting enriched KGs also increased substantially in size and association counts (see **Table 4**). These expanded graphs preserve full semantic typing, provenance annotations, and Biolink-compliant relationships, enabling new multi-hop reasoning paths (e.g., Condition → Gene → Variant → Phenotype) and improving coverage for translational research applications.

| Study | KG_MI name | KG_MI size | # associations |
|---|---|---|---|
| HTP | INCLUDE_KG_HTP_MI.rdf | 31.7 MB | 803908 |
| X01-Hakonarson | INCLUDE_KG_X01-Hakonarson_MI.rdf | 21.6 MB | 562598 |
| X01-deSmith | INCLUDE_KG_X01-deSmith_MI.rdf | 6.8 MB | 187284 |
| BRI-DSR | INCLUDE_KG_BRI-DSR_MI.rdf | 2.7 MB | 87964 |
| DSC | INCLUDE_KG_DSC_MI.rdf | 21.2 MB | 694840 |
| ABC-DS | INCLUDE_KG_ABC-DS_MI.rdf | 7.6 MB | 267477 |
| TEAM-DS | INCLUDE_KG_TEAM-DS_MI.rdf | 8.9 MB | 322660 |
| BrainPower | INCLUDE_KG_BrainPower_MI.rdf | 4.4 MB | 150651 |
| ALL (merged) | INCLUDE_ALL_MI.rdf | 64.5 MB | 1633480 |

**TABLE 4. Enriched RDF Knowledge Graph Files and Sizes**

*KNOWLEDGE DISCOVERY*

**Graph Embedding**

We trained a TransE embedding model on the merged INCLUDE KG (**ALL**) and evaluated performance using rank-based metrics. The model achieved a consistently high **Adjusted Geometric Mean Rank Index (AGMRI)**[47] across evaluation modes—**0.9985** (both, optimistic) for the combined head–tail prediction task—indicating that true triples were ranked very close to the top among all possible candidates. Similarly, the **Adjusted Arithmetic Mean Rank Index (AAMRI)**[47] was **0.9793** (both, optimistic), further reflecting strong predictive capability.

For direct ranking metrics, the **Hits@10**[48] score reached **0.305** (both, optimistic), meaning that nearly one-third of correct triples were retrieved within the top 10 predictions. Tail prediction generally outperformed head prediction (e.g., Hits@10: **0.4780** vs. **0.1324**), suggesting that the learned embeddings captured object-entity semantics more effectively than subject-entity semantics. Median rank for tail predictions was **13** compared to **1044** for head predictions, reinforcing this asymmetry.

We repeated the same training and evaluation for the **HTP** subset of the KG. While overall trends were consistent with the ALL model, the HTP embeddings exhibited higher relative performance in tail prediction tasks, reflecting the more focused and homogeneous nature of the subset graph.

We performed analyses on **both the merged KG (ALL)** and a **study-specific subgraph (HTP)** to capture complementary insights. The ALL embeddings demonstrate integration capacity—how well the model organizes and relates entities across heterogeneous datasets—while the HTP embeddings highlight fine-grained, within-study semantic structure, reducing cross-study variability and enabling focused clinical or phenotypic discovery.

To qualitatively assess embedding structure, we applied **PCA** (25 components) followed by **UMAP** and **t-SNE** to project high-dimensional embeddings into two dimensions[49]. The global embedding map for **ALL** (**Figure 4, right)** revealed distinct clustering of entities by semantic category, with participants, phenotypes, genes, and diseases forming separate, coherent regions. The **participant-only view** (**Figure 4, left**) showed strong study-based clustering, suggesting that embeddings captured study-specific context alongside shared features.

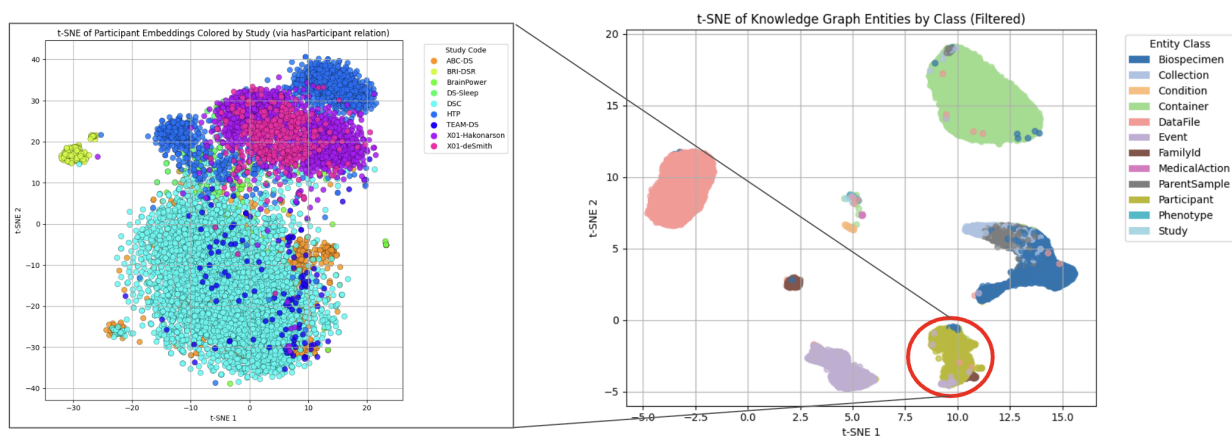

**FIGURE 4. Visualization of Knowledge Graph Embeddings Using PCA, UMAP, and t-SNE**

For **HTP** (**Figure 5**), participant embeddings colored by DS status (T21 vs. D21) showed partial separation, indicating that embeddings encode phenotype-linked variation while preserving shared attributes. In both datasets, participants occupied a cohesive subspace often proximal to relevant phenotypes and biospecimens, demonstrating that embeddings preserve both **global graph topology** and **fine-grained relationships**, enabling downstream applications such as patient stratification, outlier detection, and link prediction.

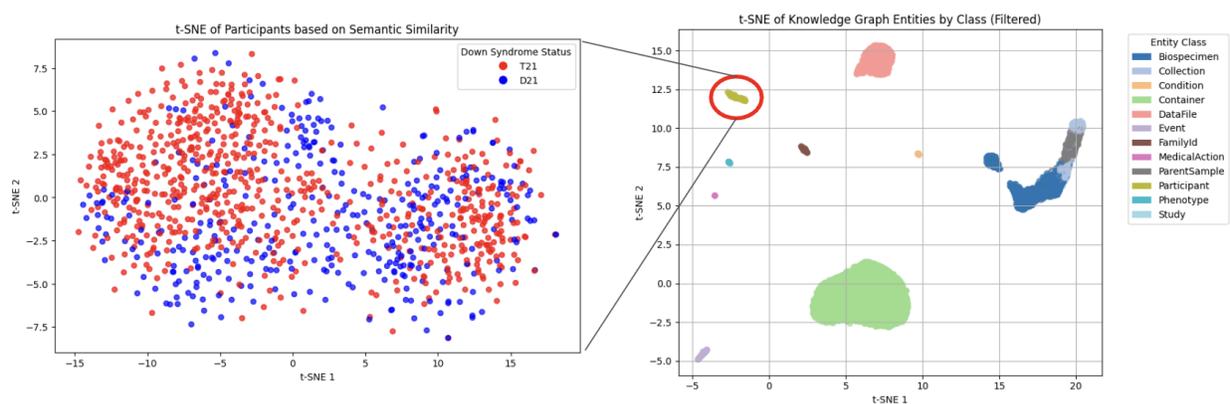

**FIGURE 5. HTP Participant Embeddings Colored by Down Syndrome Status**

To evaluate the utility of embeddings in downstream tasks, we trained a **Random Forest classifier**[50] to predict **DS status** using the entity embeddings as features. For ALL, classification achieved **92% accuracy**, with high precision and recall for T21 participants (precision: 0.93, recall: 0.98) but lower recall for D21 participants (recall: 0.50). The HTP-specific classifier showed lower overall performance (accuracy: 70%), reflecting the smaller, less balanced dataset, with stronger recall for T21 (0.92) than D21 (0.30)[51]. Confusion matrices[51] and detailed classification reports[51] are provided in **Figures 6-a,b** for ALL and HTP, respectively.

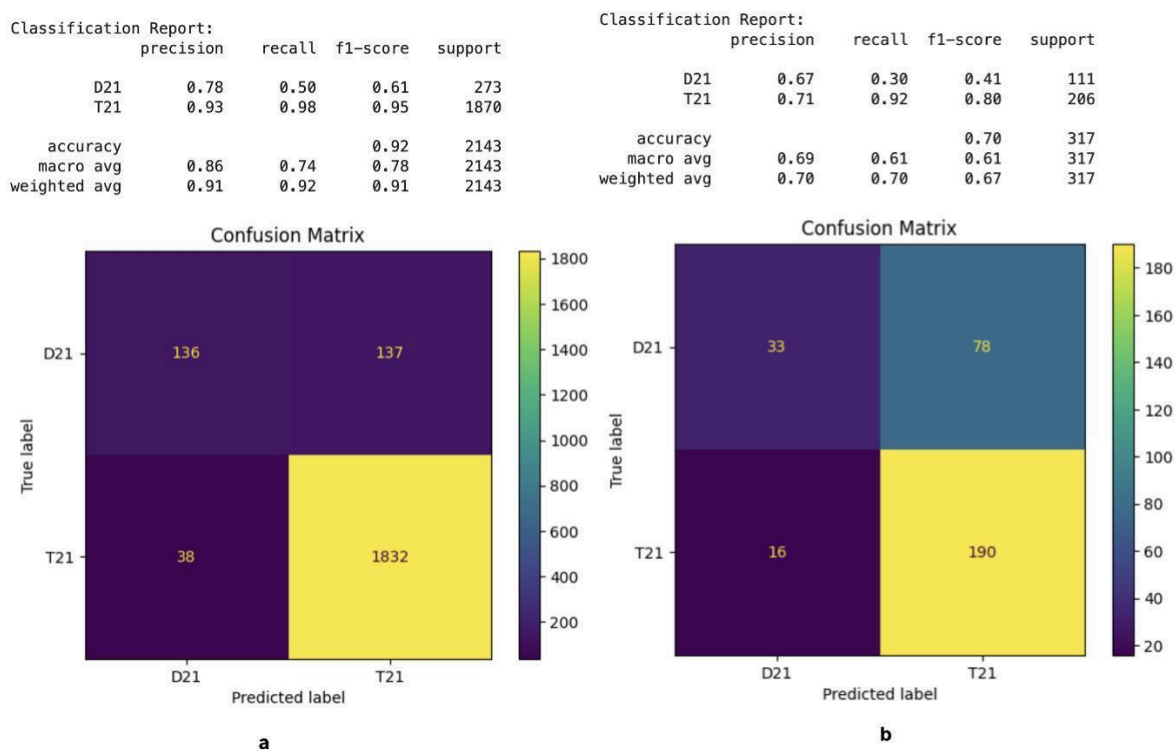

**FIGURE 6. Random Forest Classification of DS Status Using Knowledge Graph Embeddings (a)** ALL embeddings: confusion matrix and classification report (92% accuracy; strong performance for T21, lower recall for D21). **(b)** HTP embeddings: confusion matrix and classification report (70% accuracy; reduced performance due to smaller, imbalanced dataset).

**Graph Analysis**

We performed a path-based exploration of the **enriched, merged INCLUDE KG**, integrating participant-level data across nine individual studies. This approach maps gene–phenotype and gene–condition relationships while leveraging the enriched semantic links added during KG construction.

We focused on the JAK–STAT pathway[52,53] genes as a use case (JAK1, JAK2, JAK3, STAT1, STAT2, and STAT3). For each gene, all paths leading to phenotypes or conditions associated with participants were extracted from the KG using a breadth-first search approach implemented with NetworkX and rdflib.

The analysis workflow involved:

1. **Path Extraction**:
    Starting from a gene node in the enriched merged KG, all paths traversing allowed predicates (e.g., `hasPhenotype, hasCondition, biolinkDisease`) were

enumerated up to a defined maximum depth. Paths containing excluded entity types (e.g., `Participant`, `Event`) were filtered out to focus on relevant biological and clinical entities.

2. **Visualization with Sankey Plot**:
   To illustrate the flow from genes to phenotypes, a Sankey diagram was generated. Genes are depicted as source nodes, and phenotypes/conditions as target nodes. The width of each edge corresponds to the number of participants contributing to the respective path.

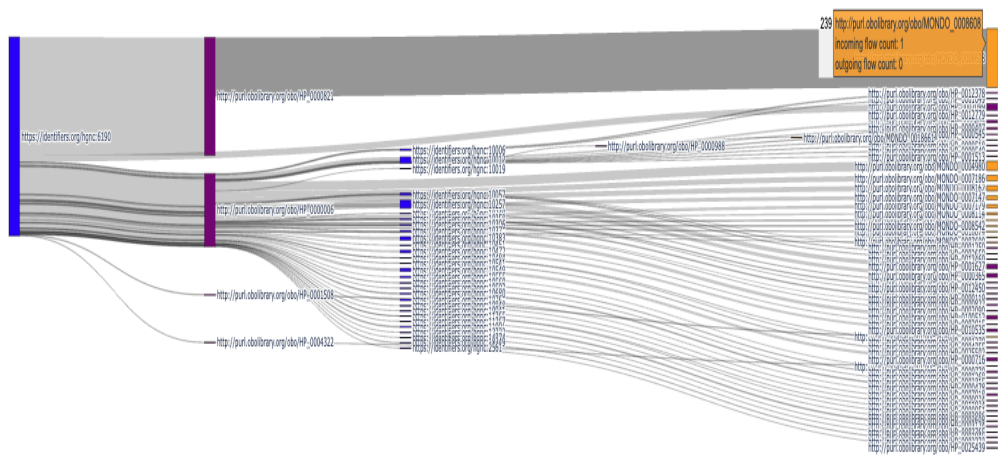

**FIGURE 7. Sankey Diagram of Gene-to-Phenotype/Condition Paths**

3. **Path Summarization and Aggregation**:
   Paths were summarized to identify unique gene–phenotype trajectories. When multiple genes were considered simultaneously, only phenotypes present in **all genes** were retained, yielding a set of 79 shared entities across the JAK–STAT gene set. A subset of these shared phenotypes includes: Hypothyroidism (`HP:0000821`), Skin rash (`HP:0000988`), Vitiligo (`HP:0001045`), Developmental delay (`HP:0001629`), etc.

4. **Comparison to Known Comorbidities**:
   Many comorbidities observed in the reference literature[53] (e.g., hypothyroidism, vitiligo, skin rash) were captured in our KG-based analysis, demonstrating the validity of the path-based approach. This method can be extended to other DS-related biological pathways and their associated gene sets, enabling systematic identification of genotype–phenotype associations.

**Interpretation**
The Sankey diagram provides an interpretable view of multi-step relationships between genes and phenotypes, complementing the high-dimensional insights obtained from graph embeddings. It highlights potential convergent effects where multiple genes influence the same phenotypic outcomes and serves as a foundation for hypothesis generation in downstream functional and clinical studies.

*KNOWLEDGE EXPLORATION*

SPARQL queries on the NIH INCLUDE KG enabled retrieval and aggregation of biomedical entities across studies. Participant-linked conditions (MONDO terms) and phenotypes (HPO terms) observed in ≥2 studies were analyzed to assess cross-study overlap. Results were summarized in tables, co-occurrence matrices, and two UpSet plots (**Figure 8** for conditions, **Figure 9** for phenotypes), highlighting patterns of overlap and uniqueness across studies.

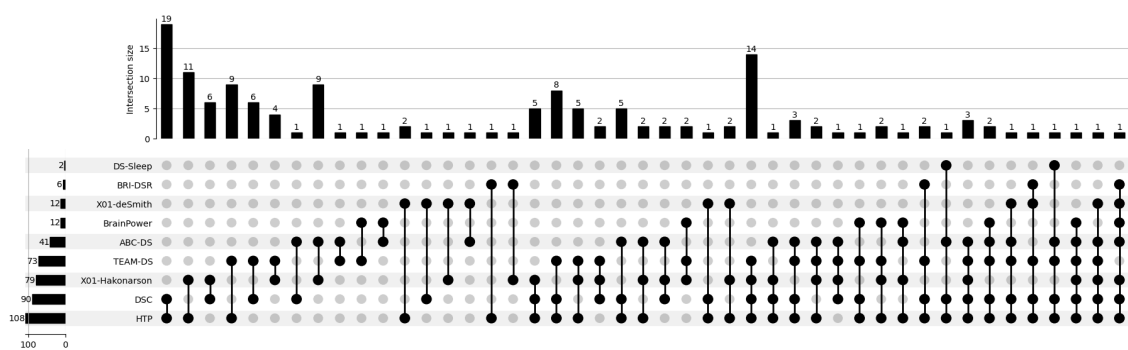

**FIGURE 8. Cross-Study Overlap of Participant-Linked Conditions (MONDO) in INCLUDE KGs**

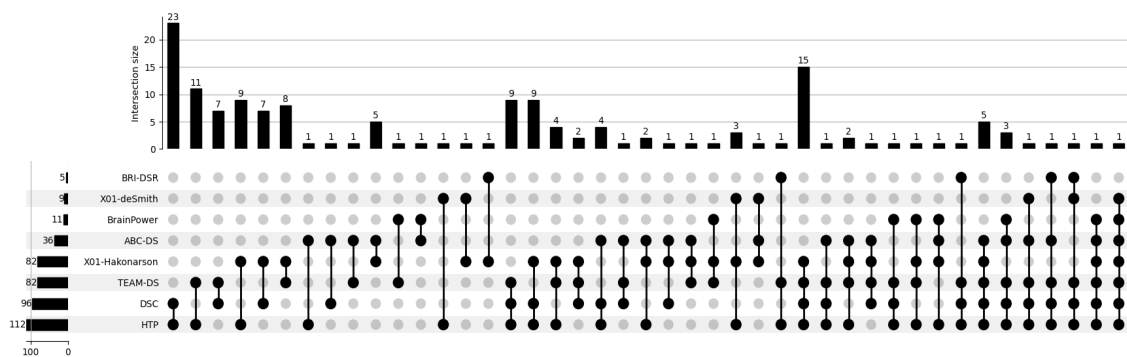

**FIGURE 9. Cross-Study Overlap of Participant-Linked Phenotypes (HPO) in INCLUDE KGs**

**DISCUSSION**

This work demonstrates the feasibility and utility of a KG–driven framework for enabling translational research in DS. By systematically transforming harmonized participant-level datasets from the NIH INCLUDE initiative into a semantically enriched, AI-ready KG, we provide an infrastructure that supports hypothesis generation, cross-study analysis, and predictive modeling in a scalable and reproducible manner.

Our framework addresses several persistent challenges in DS research. First, the heterogeneity of clinical manifestations across cohorts has historically impeded integrative analyses. By unifying data from nine distinct studies into FAIR-compliant, ontology-aligned KGs, we establish a consistent semantic foundation that enables comparative and cross-cohort exploration. Second, the incorporation of external resources through knowledge enrichment expands coverage of genes, variants, and phenotype–disease associations, creating opportunities for multi-hop reasoning and mechanistic inference that would not be possible from primary datasets alone.

The downstream analyses highlight the practical benefits of this approach. Graph embeddings achieved high performance in classifying DS status, underscoring the potential of latent graph representations for predictive modeling. Complementary path-based analyses provided interpretable, ontology-grounded evidence of shared comorbidities. Together, these findings illustrate the value of combining latent and explicit graph analytics for comprehensive discovery.

**Limitations**
The framework's utility is currently bounded by several factors.

- **Data heterogeneity and sparsity**: The richness of the KG is constrained by the completeness and quality of source data. Cohorts with limited annotation density (e.g., DS-Sleep) yielded minimal enrichment, reducing downstream predictive capacity.
- **Cohort imbalance**: Classification accuracy was influenced by unequal representation of trisomy vs. non-trisomy participants. Such imbalances may bias results and reduce generalizability. Approaches such as stratified sampling, weighting, or synthetic augmentation could improve robustness.
- **External enrichment specificity**: While external knowledge sources increased entity coverage, imported associations are not always empirically validated within INCLUDE populations, which may dilute specificity and confound interpretation.
- **Model expressivity**: The baseline embedding model (TransE) assumes translational invariance and may not fully capture hierarchical, polygenic, temporal, or context-dependent biomedical relationships. More expressive models are needed to achieve richer representations.

**Future Directions**
Several avenues exist to extend the current framework and broaden its impact:

- **Multi-omics integration**: Incorporating genomics, transcriptomics, proteomics, and metabolomics (e.g., SO, RNA-Seq, UniProt, ChEBI) would link molecular signals directly to clinical phenotypes. Such integration enables systems-level insights and cross-modal embeddings that jointly represent heterogeneous entities, positioning the KG as a discovery platform bridging omics with participant-level metadata.
- **Advanced embedding models**: Beyond TransE, approaches such as RotatE, ComplEx, DistMult, or GNNs could capture higher-order dependencies. Temporal embeddings would model disease progression and treatment effects, while hybrid architectures that combine symbolic reasoning with embeddings could improve both interpretability and predictive accuracy.
- **Integration of external knowledge**: Incorporating pathway databases, drug–target repositories, pharmacogenomic resources, and literature-derived associations would expand the scope of discoverable relationships. This integration supports cross-scale inference, drug repurposing opportunities, and discovery of novel therapeutic targets for DS comorbidities.

**Conclusion**

This work introduces a technically rigorous, extensible KG framework that transforms static harmonized datasets into dynamic discovery platforms. The demonstrated ability to support both interpretable reasoning and predictive modeling represents a methodological advance. As biomedical datasets grow in size, diversity, and complexity, such semantically constrained yet AI-compatible infrastructures will be essential for scaling precision medicine and accelerating translational discovery.

**DATA AND CODE AVAILABILITY**

In this study, we utilized harmonized participant-level datasets from the NIH INCLUDE Data Hub[54]. These datasets are accessible through the INCLUDE Data Hub as well as backend repositories such as Synapse and AWS S3, all of which require an authorized account for access. For this work, we directly accessed CSV files from Synapse, and all generated output files, including RDF representations of the KG, were stored back in Synapse with explicit links to their corresponding input CSV files to ensure full data provenance, reproducibility, and compliance with data governance requirements. The RDF files (core and enriched) are available at Synapse:syn64954214. Data provenance for both individual study-specific KGs and the merged **INCLUDE_ALL** graph has been fully captured (**Figure 10**).

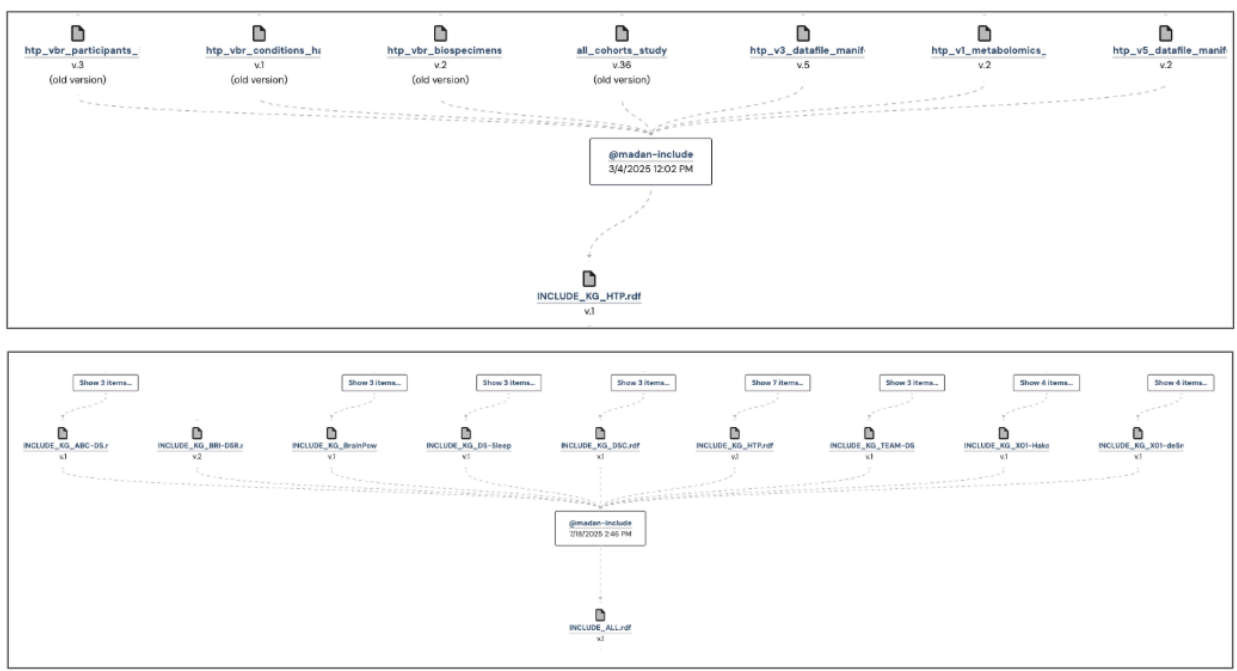

**FIGURE 10. Data Provenance and Storage Workflow for INCLUDE Knowledge Graphs**

All computational code, executables, and supporting documentation are hosted on CAVATICA[55], cloud-based data analysis platform designed for biomedical research. Access requires user registration and is available at the [INCLUDE_KG](#) repository workspace. The repository is organized according to the major framework stages—knowledge generation, knowledge enrichment, knowledge discovery, and knowledge exploration. The repository maintains a modular structure aligned with the four primary framework stages, each containing dedicated scripts and comprehensive README documentation:

**Knowledge Generation and Enrichment:**

- **KG_Schema.ipynb** and **KG_Instances.ipynb** implement the Knowledge Generation phase, where harmonized INCLUDE participant-level datasets are transformed into a FAIR-compliant RDF KG
- **KG_Enrichment_MI.ipynb** implements the Knowledge Enrichment phase, extending the schema and instance graphs with external biomedical mappings and ontologies

**Knowledge Discovery:**

- **KG_Embedding.ipynb** - KG embedding generation using TransE algorithm for learning entity representations and classification tasks leveraging embeddings for predicting participant characteristics such as DS status, vital status, and demographic features
- **KG_Analysis_Path.ipynb** - Path discovery and analysis for identifying biomedical relationships between genes, phenotypes, and participants with interactive visualizations using Sankey diagrams and UMAP dimensionality reduction for exploring entity relationships and clustering patterns

**Knowledge Exploration:**

- **KG_Chatbot.py** - Natural language query interface through an LLM-powered Streamlit chatbot application that converts user questions into SPARQL queries and displays results in interactive tables with schema-aware query generation utilizing GPT-4[56] to understand the KG structure and generate contextually appropriate queries for participant, condition, phenotype, and biospecimen data
- **KG_SPARQL.ipynb** - Multi-study condition mapping and overlap analysis to identify shared conditions across different research studies with UpSet plot visualizations showing study combination patterns and revealing collaboration opportunities between research groups

**Note**: Gaining access to the INCLUDE Data Hub does **not** provide access to Synapse or CAVATICA. Likewise, creating an account in Synapse or CAVATICA does **not** make data or code automatically available. To obtain the most current RDF files generated from INCLUDE datasets, or to access the code within CAVATICA used for generation and analysis of later versions, please reach out directly to an INCLUDE DCC team member.

**ACKNOWLEDGMENTS**

The RDF files for the knowledge graph were generated from data accessed through the INCLUDE (INvestigation of Co-occurring conditions across the Lifespan to Understand Down syndromE) Data Hub[15], which is supported by National Heart, Lung and Blood Institute grant U2CHL156291.